# Atmospheric Centers of Action: current features and expected changes from simulations with CMIP5 and CMIP6 models

I. I. Mokhov [1,2]*, A. M. Osipov [2], A. V. Chernokulsky [1]

[1]A.M. Obukhov Institute of Atmospheric Physics RAS

[2]Lomonosov Moscow State University

* e-mail: mokhov@ifaran.ru

## Abstract

The results of an analysis of changes in the characteristics of atmospheric centers of action (ACAs) in the Northern (NH) and Southern (SH) hemispheres using results of simulations with the CMIP5 and CMIP6 ensembles of climate models are presented. The ability of models to simulate ACA features is estimated for the historical scenario in comparison with ERA5 reanalysis data. The projected changes are evaluated under RCP8.5 and SSP5-8.5 scenarios for CMIP5 and CMIP6 models, respectively. The ACA intensity is evaluated that defined as the difference in sea level pressure averaged over the ACA region and the entire hemisphere. In NH, reanalysis and models show greater intensity of subtropical oceanic anticyclonic ACAs in summer than in winter. The opposite is found for the intensity of NH subpolar oceanic cyclonic ACAs. The interannual variability of the ACA intensity in winter is generally greater than in summer. In SH, the season with greater intensity of oceanic anticyclonic and cyclonic ACAs and its interannual variability varies from ocean to ocean. CMIP5 and CMIP6 models show substantial changes of ACAs characteristics in the XXI century. More significant trends in the strengthening of ACAs in the 21st century appear in the SH, especially in the winter seasons. The most consistent weakening trends are found over continents for winter North American maximum and the summer Asian minimum. For the winter Siberian maximum, the weakening trend is found more pronounced in CMIP6 models than in CMIP5.

## 1. Introduction

Atmospheric centers of action (ACAs) are large-scale structural formations in the atmosphere of the Earth's climate system that characterize the general circulation in the troposphere of the Northern (NH) and Southern (SH) hemispheres and depend on the distribution of oceans and continents. In a variable atmosphere, ACAs appear quite stable in the surface pressure field with monthly and seasonal averaging. In the atmosphere, there are quasi-permanent subpolar cyclonic regions with reduced surface pressure and subtropical anticyclonic regions with increased surface pressure over the oceans, as well as anticyclonic regions with increased surface pressure over the polar latitudes (Intense Atmospheric Vortices..., 2018).

These features are associated with the three-cell meridional circulation in the troposphere of each hemisphere with the formation of tropical Hadley cells, reverse Ferrel cells in mid-latitudes and polar cells. With a three-cell meridional circulation in the troposphere of the NH and SH, latitudinal belts with reduced surface pressure appear in the subpolar regions of the ascending branches of the Ferrel cells and polar cells and with increased surface pressure in the regions of the descending branches of the Hadley and Ferrel cells — in subtropical latitudes, as well as in polar latitudes with descending branches of polar cells. This is particularly evident in the oceanic subantarctic latitudes of the SH. The presence of continents disrupts the zonal structure of the atmospheric circulation, and due to the difference in the heat capacity of the active layers of the continents and oceans in the annual cycle, the temperature difference between them changes sign. Thus, in winter the surface of continents is colder, and in summer it is warmer than the surface of oceans at the same latitudes. At the same time, seasonal ACAs of different vorticity are formed over the continents, i.e., cyclonic in summer and anticyclonic in winter (Intense Atmospheric Vortices..., 2018).

In the NH, in the pressure field over the oceans, the Aleutian and Icelandic cyclonic ACAs in subarctic latitudes and the Azores and Hawaiian anticyclonic ACAs in subtropical latitudes are distinguished. In polar latitudes, the Arctic and the Greenland highs occur. Over the continents, anticyclonic ACAs — the Siberian and North American highs — appear in the winter seasons, while cyclonic ACAs — the Asian and North American lows — in the summer seasons (Intense Atmospheric Vortices..., 2018).

In the SH, anticyclonic ACAs — South Atlantic, South Indian (Mascarene), and South Pacific highs — form over the oceans at subtropical latitudes. In the subantarctic oceanic latitudes, there is a continuous low-pressure zone, with the greatest decrease in pressure in the areas of the Indian Ocean, South Atlantic, and South Pacific lows. Seasonal ACAs occur over the continents, including the South African, Australian and South American lows in summer. In the polar latitudes, the Antarctic highs appears (Intense Atmospheric Vortices..., 2018).

Characteristic regional features as well as key processes on a hemispheric and global scale are associated with the ACAs (Climate Change 2013; Climate Change 2021). In the area of influence of the Siberian High, the lowest near-surface temperature in the Northern Hemisphere observed during the winter months. A number of important indicators of variations in the general circulation of the atmosphere are directly determined by the characteristics of the corresponding ACA. In particular, the variations of the North Atlantic Oscillation are determined by the intensity of the Icelandic Low and Azores High. In the zone of influence of the Siberian High in the northern part of Asia, an area with the largest near-surface warming in recent decades during the winter season has been identified (Intense Atmospheric Vortices..., 2018).

Many studies are devoted to the analysis of the nature of ACAs (Rossby, 1939; Haurwitz, 1940; Blinova, 1943; Smagorinsky, 1958; Wallace, 1988; Galin and Kharitonenko, 1989; Perevedentsev et al., 1994; Gushchina and Petrosyants, 1998; Mokhov and Petukhov, 2000; Cohen et al., 2001; Mokhov and Khon, 2005; Khon and Mokhov; 2006; Zheleznova and Gushchina, 2016; Sun et al., 2017; Intense Atmospheric Vortices..., 2018; Mokhov et al., 2020; Mokhov et al., 2021; Mokhov et al., 2022). In (Rossby, 1939; Haurwitz, 1940; Blinova, 1943) modes of ACA formation were studied using Rossby-Blinova waves. A significant part of the energy of these planetary waves is associated with the stationary component, which manifests itself in positive and negative anomalies of the time-averaged pressure field in the troposphere. On the other hand, ACAs can be characterized as large-scale vortex structures (Intense Atmospheric Vortices..., 2018). In (Galin and Kharitonenko, 1989) it was concluded that large-scale anomalies of the pressure field in the lower troposphere are mainly determined by the thermal factor, although their localization and intensity also depend on the orographic influence. In (Mokhov and Petukhov, 2000), analytical expressions were obtained for the mode that makes the main contribution to the formation of the ACA. The model expressions obtained in (Mokhov and Petukhov, 2000) allow for a qualitative analysis of the sensitivity of the ACA characteristics to global anthropogenic changes caused by changes in the content of greenhouse gases in the atmosphere, and to natural variability associated, for example, with phenomena such as El Niño — Southern Oscillation. Global climate changes that affect vortex activity in the atmosphere are also manifested in the features of the ACAs. Significant near-surface warming has been observed in recent decades for the winter season in the area of influence of the Siberian anticyclone. Significant surface temperature anomalies in the northern part of North America are associated with the regime of the Aleutian cyclone (Gushchina and Petrosyants, 1998; Trenberth, 1998; Mokhov and Petukhov, 2000; Mokhov and Khon, 2005). Significant large-scale anomalies in atmospheric variables such as temperature, precipitation, and cloudiness are associated with the ACAs.

In general, according to observational and reanalysis data, there is a significant variability in the ACAs characteristics (Intense Atmospheric Vortices..., 2018). In recent decades, statistically significant long-term trends for the ACAs characteristics have been noted under the global-scale temperature changes, especially in winter in the NH (Mokhov and Petukhov, 2000; Mokhov and Khon, 2005). The obtained estimates indicate possible significant changes in the central pressure with continued global warming. The use of modern climate models makes it possible to assess changes in the characteristics of the ACA under expected global climate changes (Intense Atmospheric Vortices..., 2018; Climate Change 2013; Climate Change 2021). In (Khon and Mokhov, 2006), estimates of the sensitivity of changes in ACA characteristics to changes in hemispheric surface temperature were obtained using the results of numerical calculations with various climate models. In particular, Khon and Mokhov (2006) found a tendency for the Icelandic ACA to increase under warming in the 21st century according to model simulations. A weakening of the continental winter ACA was found. For the Pacific ACAs, non-linear changes were found for the 21st century (Mokhov et al., 2022).

In (Intense Atmospheric Vortices..., 2018; Mokhov et al., 2021), estimates of possible changes in the ACAs characteristics in the 21st century under various RCP scenarios of anthropogenic forcings are presented using the results of numerical calculations with climate models of the CMIP5 family. A general deepening of the cyclonic ACA was found in both hemispheres by the end of the 21st century compared to current conditions. At the same time, different trends were noted for the North American Low formed during warm seasons from simulations with different models. Over the continents in the NH, a general weakening of the winter anticyclonic ACAs is observed. The noted increase in the intensity of oceanic subtropical maxima, in particular in the SH, can be associated with changes in the meridional Hadley cell under global warming. Different trends were estimated for the subtropical Azores and Hawaiian high in the NH. The noted weakening of the polar ACAs is more pronounced in cold seasons, which may be associated with a shift in the trajectories of extratropical cyclones to high latitudes. A more pronounced weakening was found under the scenario of stronger anthropogenic forcings (RCP8.5) for the 21st century (Mokhov et al., 2022).

In (Mokhov et al., 2022) the results of an analysis of possible changes in the characteristics of ACAs in the Northern Hemisphere by simulations with the CMIP5 and CMIP6 ensembles of climate models under scenarios of anthropogenic impacts RCP8.5 and SSP5-8.5 in the 21st century are presented. Quite consistent estimates from simulations with climate models of the CMIP6 and CMIP5 ensembles were obtained for the weakening trends of the winter North American High and the summer Asian Low. For the winter Siberian High, the weakening trend was found to be more significant by simulations with the CMIP6 ensemble of climate models. In a comparative analysis of the ACAs detected by model simulations and reanalysis data, it was found that, in general, the intensity of subtropical anticyclonic ACAs over the Atlantic and Pacific Oceans is greater for the summer seasons than for the winter seasons according to reanalysis data and model simulations. The opposite is found for the intensity of subpolar cyclonic ACAs over the Atlantic and Pacific oceans. Their intensity is generally greater in winter than in summer. At the same time, the interannual variability of ACA intensity is also greater in winter than in summer.

This paper provides a comprehensive and detailed analysis of the current features and expected changes in the 21st century of the major ACAs in both hemispheres, based on simulations with the CMIP5 and CMIP6 ensembles of climate models. For the ACA of the Southern Hemisphere, such assessments have not been carried out before.

## 2. Data and analysis methods used

To analyze the ACAs, we used the results of simulations of pressure fields at sea level in the NH and SH with ensembles of climate models from the CMIP5 and CMIP6 projects in comparison with reanalysis data. In particular, the results of numerical calculations with climate models of the CMIP5 and CMIP6 ensembles were analyzed under the "Historical" scenario and under the RCP8.5 and SSP5-8.5 scenarios for the 21st century. The quality of simulations of the ACAs characteristics ftom model simulations was assessed in comparison with the ERA5 reanalysis data (Hersbach et al., 2020) for the base period 1981-2005.

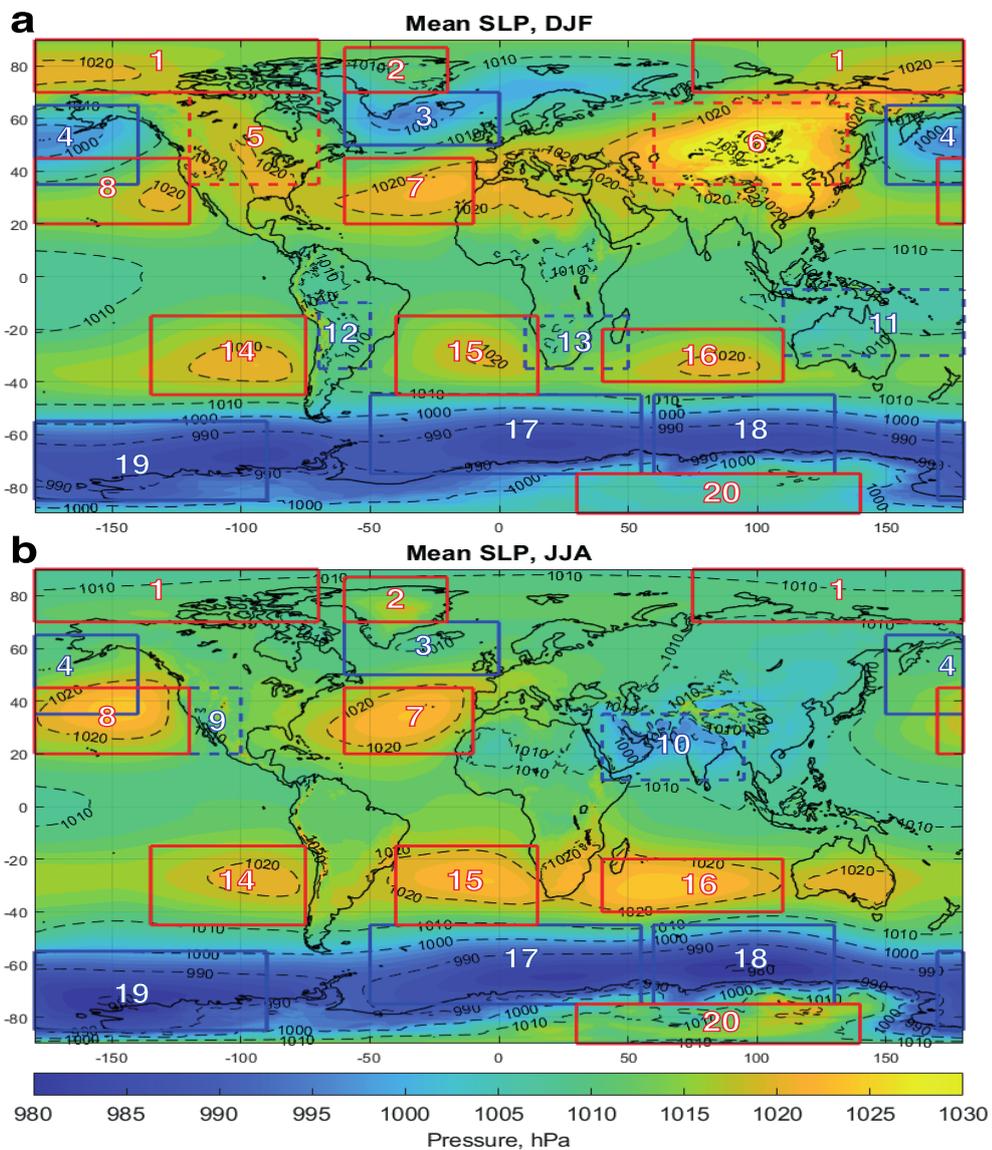

Figure 1. The location of the ACAs analyzed in the paper, in December-January-February (top) and June-July-August (bottom). The red outline shows the maxima, the blue outline shows the minima. The numbers correspond to those in Table 1.

The areas corresponding to each of the analyzed ACAs were identified similarly to (Intense Atmospheric Vortices…, 2018, Mokhov et al., 2020), including 10 ACAs in the NH and 10 ACAs in the SH (Fig. 1, Table 1). This includes 11 ACAs over the oceans, 9 ACAs over land, 13 year-round ACAs and 7 seasonal ACAs. Other ACAs also appear to a weaker extent, such as the maximum over Australia in winter (Fig. 1), but are not analyzed in this study.

Table 1. Characteristics of the analyzed atmospheric centers of action

| № | ACAs | seasons | surface type | region | |
|---|------|---------|--------------|--------|---|
|   |      |         |              | latitude | longitude |
| Northern Hemisphere (NH) | | | | | |
| 1 | Arctic High | year-round | ocean/ice | 70–90°N | 75°E – 70°W |
| 2 | Greenland High | year-round | land/ice | 70–87°N | 20–60°W |
| 3 | Islandic Low | year-round | ocean | 50–70°N | 0–60°W |
| 4 | Aleutian Low | year-round | ocean | 35–65°N | 150°E – 140°W |
| 5 | North American High | winter NH | land | 35–70°N | 70–120°W |
| 6 | Siberian High | winter NH | land | 35–66°N | 60–135°E |
| 7 | Azores High | year-round | океан | 20–45 N | 10–60°W |
| 8 | Hawaiian High | year-round | океан | 20–45°N | 170°E – 120°W |
| 9 | North American Low | summer NH | land | 20–45°N | 100–120°W |
| 10 | Asian Low | summer NH | land | 10–35°N | 40–95°E |
| Southern Hemisphere (SH) | | | | | |
| 11 | Australian Low | summer SH | land | 5–30°S | 110–180°E |
| 12 | South American Low | summer SH | land | 10–35 S | 50–70°W |
| 13 | South African Low | summer SH | land | 15–35°S | 10–50°E |
| 14 | South Pacific High | year-round | ocean | 15–45°S | 75–135°W |
| 15 | South Atlantic High | year-round | ocean | 15–45°S | 40°W – 15°E |
| 16 | South Indian High | year-round | ocean | 20–40°S | 40–110°E |
| 17 | South Atlantic Low | year-round | ocean | 45–75°S | 50°W – 55°E |
| 18 | Indian Ocean Low | year-round | ocean | 45–75°S | 60–130°E |
| 19 | South Pacific Low | year-round | ocean | 55–85°S | 90–170°W |
| 20 | Antarctic High | year-round | land/ice | 75–90°S | 30–140°E |

The analysis determined the average value of sea level pressure $P_c$ in areas of anticyclonic or cyclonic central air pressure - respectively, with increased or decreased pressure relative to background conditions. In particular, seasonal averages were analyzed, in particular for the winter and summer seasons. The ACA intensity was characterized by the sea level pressure in the ACA region $P_c$ (similar to (Mokhov et al., 2020)) and the corresponding pressure difference $I_c$ relative to the hemispheric mean sea level pressure $P_H$:

$$I_c = P_c - P_H. \qquad (1)$$

Relative changes in the ACA intensity $I_c' = I_c/\delta I_c$ were assessed by normalizing $I_c$ to the corresponding mean for the 1981–2005 base period $\delta I_c$. Other characteristics of the ACA were also analyzed similarly to (Mokhov et al., 2020), including the values of the absolute maximum (minimum) pressure in the center of the ACA $I_{cm}$ and its latitude $\varphi_C$ and longitude $\lambda_C$ (see also (Mokhov et al., 2021; Mokhov et al., 2022a,b; Mokhov et al., 2023).

To analyze the characteristics of the ACA, results of simulations with climate models from the CMIP5 and CMIP6 projects were utilized. The selection of models was based on the presence of a single set of climate scenarios, including for the future climate, as well as the uniformity of the global attributes of the model run, including realisation (ensemble member number), implementation method, physics and forcing. In this study, only models with variants coded as 'r1i1p1' and 'r1i1p1f1', were selected from all CMIP5 and CMIP6 models, respectively. A total of 25 models were selected for the CMIP5 project (Table 1) and 32 models for the CMIP6 project (Table 2). Ensembles with these models are hereafter denoted as "all" models.

Table 2. Climate models used in this study

| № | CMIP6 models | | CMIP5 models | |
|---|---|---|---|---|
| | models | model resolution (longitude x latitude) | models | model resolution (longitude x latitude) |
| 1 | ACCESS-CM2 | 1.875° x 1.25° | ACCESS1-0 | 1.875° x 1.25° |
| 2 | ACCESS-ESM1-5 | 1.875° x 1.25° | ACCESS1-3 | 1.875° x 1.25° |
| 3 | AWI-CM-1-1-MR | 0.9375° x 0.9375° | BCC-CSM1-1 | 2.8125° x 2.8125° |
| 4 | BCC-CSM2-MR | 1.125° x 1.125° | BNU-ESM | 2.8° x 2.8° |
| 5 | CAMS-CSM1-0 | 1.125° x 1.125° | CanESM2 | 2.8125° x 2.76° |
| 6 | CanESM5 | 2.8° x 2.8° | CCSM4 | 1.25° x 0.94° |
| 7 | CAS-ESM2-0 | 1.4° x 1.39° | CESM1-BGC | 1.25° x 0.94° |
| 8 | CESM2-WACCM | 1.25° x 0.94° | CESM1-CAM5 | 1.25° x 0.94° |
| 9 | CIESM | 1.25° x 0.94° | CMCC-CM | 0.75° x 0.75° |
| 10 | CMCC-CM2-SR5 | 1.25° x 0.94° | CMCC-CMS | 1.88° x 1.86° |
| 11 | CMCC-ESM2 | 1.25° x 0.94° | CNRM-CM5 | 1.4° x 1.39° |
| 12 | EC-Earth3 | 0.7° x 0.7° | FIO-ESM | 2.8° x 2.8° |
| 13 | EC-Earth3-Veg | 0.7° x 0.7° | GISS-E2-H | 2.5° x 2° |
| 14 | EC-Earth3-Veg-LR | 1.125° x 1.125° | GISS-E2-R | 2.5° x 2° |
| 15 | FGOALS-f3-L | 1.25° x 1° | INM-CM4 | 2° x 1.5° |
| 16 | FGOALS-g3 | 2° x 2.25° | IPSL-CM5A-LR | 3.75° x 1.9° |
| 17 | FIO-ESM-2-0 | 1.25° x 0.9375° | IPSL-CM5A-MR | 2.5° x 1.2676° |
| 18 | GFDL-ESM4 | 1.25° x 1° | MIROC5 | 1.4° x 1.39° |
| 19 | IITM-ESM | 1.875° x 1.904° | MIROC-ESM | 2.8125° x 2.79° |
| 20 | INM-CM4-8 | 2° x 1.5° | MIROC-ESM-CHEM | 2.8125° x 2.79° |
| 21 | INM-CM5-0 | 2° x 1.5° | MPI-ESM-LR | 1.875° x 1.85° |
| 22 | IPSL-CM6A-LR | 2.5° x 1.25° | MPI-ESM-MR | 1.875° x 1.85° |
| 23 | KACE-1-0-G | 1.875° x 1.25° | MRI-CGCM3 | 1.125° x 1.1° |
| 24 | KIOST-ESM | 1.88° x 1.86° | NorESM1-M | 2.5° x 1.9° |
| 25 | MIROC6 | 1.40625° x 1.40625° | NorESM1-ME | 2.5° x 1.9° |
| 26 | MPI-ESM1-2-HR | 0.9375 x 0.935° | | |
| 27 | MPI-ESM1-2-LR | 1.875° x 1.8652° | | |
| 28 | MRI-ESM2-0 | 1.125° x 1.125° | | |
| 29 | NESM3 | 1.875° x 1.865° | | |
| 30 | NorESM2-LM | 2.5° x 1.89474° | | |
| 31 | NorESM2-MM | 1.25° x 0.9375° | | |
| 32 | TaiESM1 | 1.25° x 0.9375° | | |

For further analysis, the climate models that most adequately simulated the ACAs (so-called "best" models) were identified using two criteria. The first criterion is based on the degree of adequacy of a model to simulate the global surface atmospheric pressure field, which is estimated by comparing the modeled pressure fields (under the historical scenario) with those from the ERA5 reanalysis data for the base period 1981-2005 for different seasons. The second criterion takes into account the number of ACAs whose $I_c$ values simulated by a model fit the $I_c$ standard deviation obtained by ERA5. A model was included in the sample of "best" models if at least one of the two criteria placed it in the top quarter of "all" models and the other criterion placed it in the top half.

## 3. Results

### 3.1 Determining the "best" models

Figure 2 shows Taylor diagrams assessing the degree of agreement between model estimates of sea-level pressure field in the NH in different climate models and the corresponding estimates from the ERA5 reanalysis data for the base period 1981-2005: by simulations with the CMIP5 (a, b) and CMIP6 (c, d) ensembles of climate models for winter (a, c) and summer (b, d) seasons. Figure 3 shows the corresponding values for the SH.

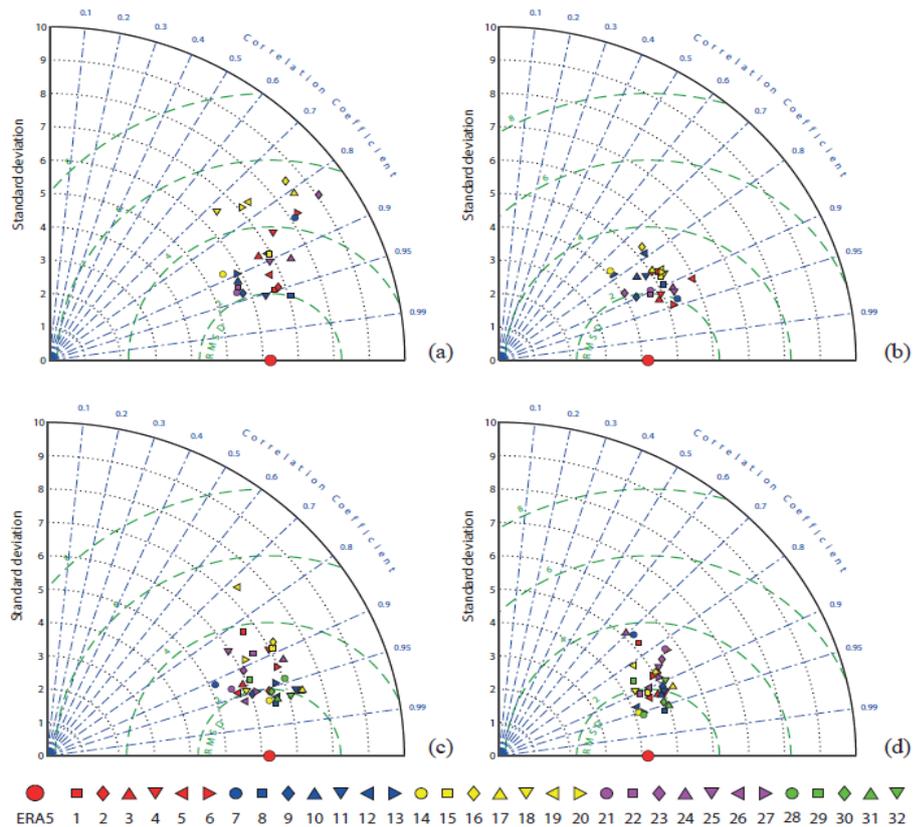

Figure 2. Taylor diagrams characterizing the degree of correspondence of the atmospheric pressure field at sea level in the NH from simulations with full ensembles of the climate models CMIP5 (a, b) and CMIP6 (c, d) ("historical" scenario) and according to ERA5 reanalysis data for the winter (DJF) (a, c) and summer (JJA) (b, d) seasons for the base period 1981-2005. The radial coordinate characterizes the spatial standard deviation of the pressure (hPa), the angular coordinate is the coefficient of spatial correlation of the pressure field between the results of the model calculations and the reanalysis data. The green dotted line shows the standard deviation (in hPa) of the results

of model calculations relative to the corresponding estimates from the reanalysis data. Model numbers are the same as in Table 2.

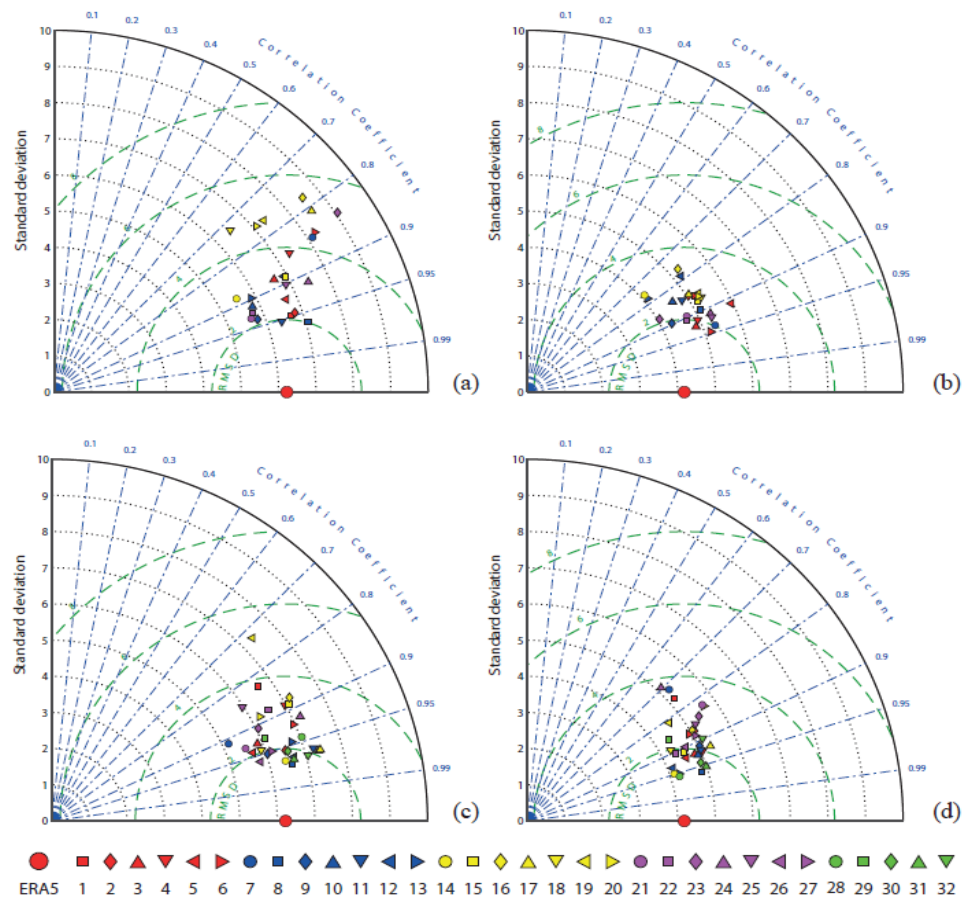

Figure 3. The same as in Fig. 2, but for the SH: (a, c) for the winter (JJA) season, (b, d) for the summer (DJF) season. Model numbers correspond to Table 2.

According to Fig. 2, simulations of the sea level pressure field are generally better in the CMIP6 climate models than in the CMIP5 climate models. The results for different seasons by simulations with the CMIP5 ensemble of climate models indicate, in general, a noticeably better reproduction of the atmospheric pressure field in winter (Fig. 2a) than in summer (Fig. 2b). At the same time, the reproduction of the atmospheric pressure by simulations with the CMIP6 climate models is significantly better than with the CMIP5 climate models, both in winter and in summer. In the SH, the general agreement of model simulations of the atmospheric pressure field (including the ACAs intensity) with the ERA5 reanalysis is better for the CMIP6 climate models than for CMIP5 climate models (Fig.3), but not as significant as for the NH.

In general, in the NH, climate models reproduce the ACAs better in winter, while such a dependence does not appear for the SH (Figures. 4, 5, Table 3). In particular, a number of models in both the CMIP5 and CMIP6 ensembles successfully reproduce all ACAs in the NH winter. In the NH summer, a number of models do not reproduce any of the ACAs. In the SH, there is no single model that reproduces all ACAs, but there is no single model that does not reproduce a single ACA (both in winter and in summer, both for CMIP5 and CMIP6 models). There was a slight improvement in the reproduction of the ACAs by the CMIP6 models compared to the CMIP5 models in both hemispheres in both winter and summer (Table 3), but for some ACAs a deterioration in the reproduction was noted in CMIP6 compared to CMIP5 (e.g., for the Mascarene and South Pacific Highs) (Fig. 6). In general, year-round ACAs are better reproduced than seasonal ones (Fig. 6). Among all ACAs, the South Pacific High (for both JJA and DJF) and the Icelandic Low (for DJF) are the best reproduced, while the South American Low and the Antarctic High are the worst reproduced (Fig. 6).

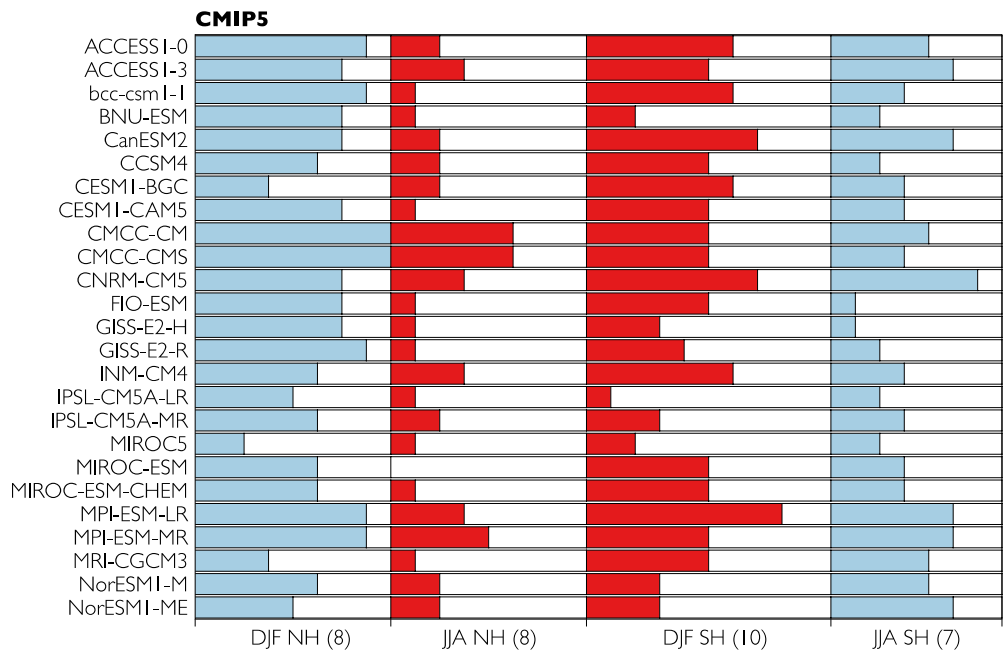

Figure 4. The number of ACAs for which the agreement of $I_c$ values (within their standard deviation) was found according to the ERA5 reanalysis data and according to the model simulations for the corresponding CMIP5 models for different hemispheres and different seasons. The total number of ACAs for each hemisphere/season is given in brackets below. ACAs in the winter season are shown in blue, in the summer season in red.

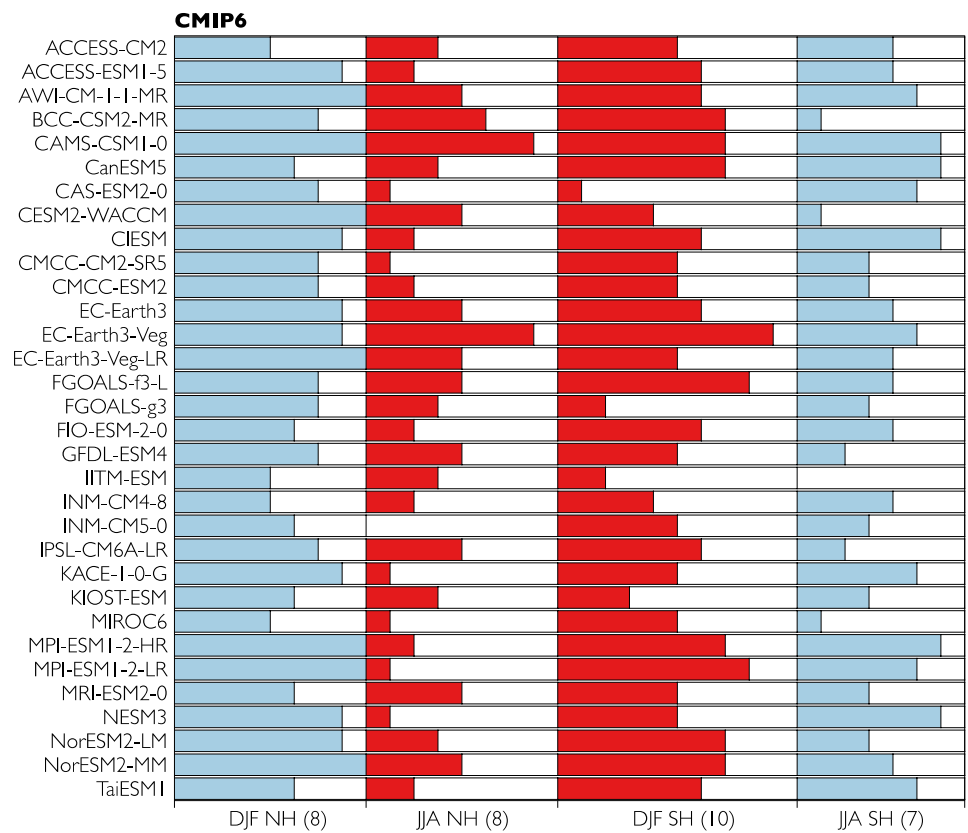

Figure 5. Same as Figure 4, but for CMIP6 models.

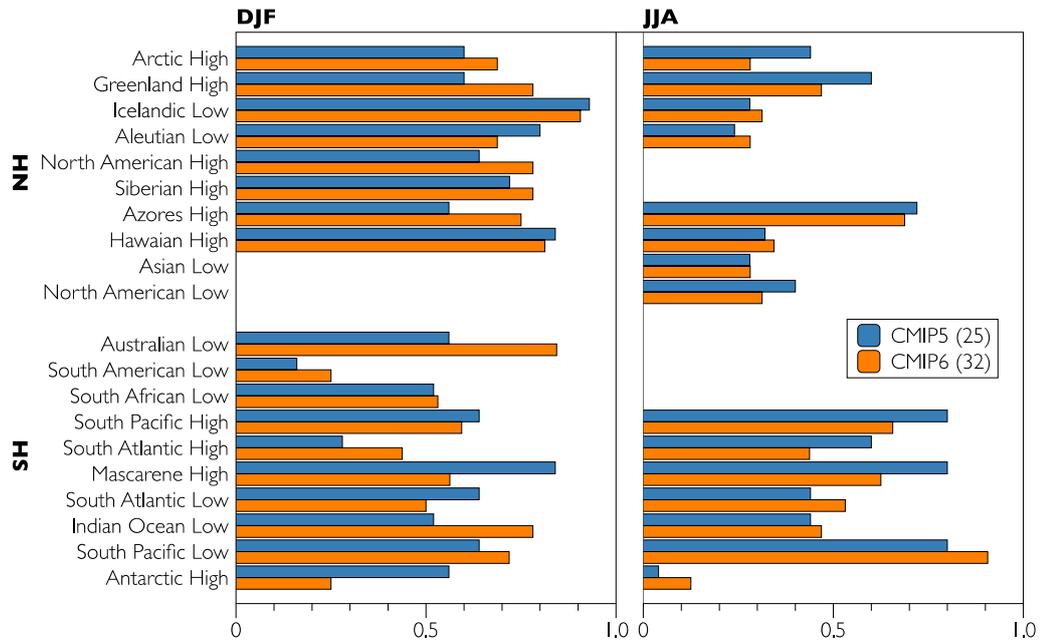

Figure 6. Fraction of models for which the agreement of $I_c$ values (within their standard deviation) with the corresponding value from the ERA5 reanalysis data is found for different ACAs for different seasons (the total number of models for each ensemble is given in parentheses in the legend).

Table 3. Average number of successfully reproduced ACAs in the ensemble of models (and as a percentage of the total number of ACAs)

|  | NH, DJF (8) | NH, JJA (8) | SH, DJF (10) | SH, JJA (7) |
|---|---|---|---|---|
| CMIP5 | 5.6 (70%) | 2.0 (25%) | 4.7 (47%) | 3.3 (47%) |
| CMIP6 | 6.2 (78%) | 2.9 (36%) | 5.5 (55%) | 3.8 (54%) |

Based on the results of a comparative analysis with reanalysis data, 9 models of the CMIP6 ensemble were selected (AWI-CM-1-1-MR, CAMS-CSM1-0, CIESM, EC-Earth3, EC-Earth3-Veg, FGOALS-f3-L, MPI-ESM1-2-HR, MPI-ESM1-2-LR, TaiESM1) and 8 CMIP5 ensemble models (ACCESS1-0, ACCESS1-3, BCC-CSM1-1, CESM1-CAM5, CMCC-CM, CNRM-CM5, MPI -ESM-LR, NorESM1-ME), which best reproduce the regimes of the analyzed ACAs for the base period in comparison with the reanalysis data.

## 3.2 Comparison of ACA characteristics from model simulations and reanalysis data for current climate

### 3.2.1. Northern Hemisphere

Table 4 shows quantitative estimates of the intensity $I_c$ of the key ACAs in the NH for the winter and summer seasons estimated by the ERA5 reanalysis data and from the ensemble means by simulations with "all" and "best" CMIP5 and CMIP6 models for the base period 1981-2005. The interannual standard deviations are also presented. The seasonal intensity of ACAs cannot always be significantly determined against the background of their interannual variability, especially when it is determined according to (1) as the average pressure anomaly for a large area. The modes of interdecadal climate variability that influence the characteristics of the ACAs, are also should be taken into account. In particular, the North Atlantic Oscillation is associated with the Icelandic Low and the Azores High.

Table 4. Intensity $I_c$ [hPa] of the key ACAs in the NH for the winter (a) and summer (b) seasons estimated by the ERA5 reanalysis data and from the ensemble means by simulations with "all" CMIP5 and CMIP6 models for the base period 1981-2005. The interannual standard deviations of the ACA intensity are given in parentheses. The corresponding estimates for the "best" models are given in square brackets.

(a)

| ACAs Northern Hemisphere | winter (December-January-February) | | |
|---|---|---|---|
| | reanalysis | CMIP5 models | CMIP6 models |
| Arctic High | 2.2 (±3.4) | 0.9 (±3.6) [2.2 (±3.3)] | 2.3 (±3.3) [1.9 (±3.5)] |
| Greenland High | -5.0 (±3.9) | -3.3 (±4.3) [-4.1 (±3.7)] | -4.3 (±4.2) [-5.0 (±4.5)] |
| Islandic Low | -10.7 (±3.0) | -11.3 (±2.8) [-11.2 (±2.8)] | -11.0 (±3.0) [-11.7 (±3.2)] |
| Aleutian Low | -9.6 (±2.7) | -9.5 (±3.2) [-8.5 (±3.1)] | -9.5 (±3.3) [-9.5 (±3.4)] |
| North American High | 3.1 (±1.0) | 2.6 (±1.3) [2.9 (±1.3)] | 3.0 (±1.2) [3.2 (±1.3)] |
| Siberian High | 10.6 (±1.2) | 10.4 (±1.4) [11.0 (±1.3)] | 10.8 (±1.5) [10.6 (±1.5)] |
| Azores High | 4.9 (±1.6) | 5.2 (±1.6) [5.3 (±1.5)] | 4.9 (±1.8) [5.0 (±1.8)] |
| Hawaiian High | 0.1 (±2.0) | -0.1 (±2.1) [0.2 (±2.1)] | -0.6 (±2.3) [-0.7 (±2.3)] |

(b)

| ACAs Northern Hemisphere | summer (June-July-August) | | |
|---|---|---|---|
| | реанализ | модели CMIP5 | модели CMIP6 |
| Arctic High | -0.8 (±1.6) | -1.8 (±1.9) [-1.8 (±1.7)] | -1.4 (±1.7) [-1.8 (±1.7)] |
| Greenland High | 1.5 (±2.0) | 2.6 (±2.5) [1.7 (±2.4)] | 0.2 (±2.3) [1.6 (±2.2)] |
| Islandic Low | 0.1 (±1.1) | 0.0 (±1.2) [-0.6 (±1.2)] | -1.0 (±1.3) [-0.8 (±1.3)] |
| Aleutian Low | 3.4 (±0.7) | 2.1 (±1.1) [2.2 (±1.0)] | 2.2 (±1.1) [2.1 (±1.2)] |
| Azores High | 8.2 (±0.6) | 5.8 (±0.7) [5.8 (±0.8)] | 5.8 (±0.8) [5.9 (±0.7)] |
| Hawaiian High | 7.5 (±0.6) | 5.6 (±0.8) [5.3 (±0.8)] | 5.5 (±0.8) [5.4 (±0.8)] |
| North American Low | 0.9 (±0.6) | -1.8 (±0.6) [-2.8 (±0.6)] | -3.1 (±0.7) [-2.2 (±0.6)] |
| Asian Low | -8.4 (±0.3) | -11.2 (±0.7) [-12.0 (±0.7)] | -11.7 (±0.7) [-11.6 (±0.6)] |

For the ACAs in Table 4a, significant (at the 99% or higher level of statistical significance) estimates of the mean winter intensity $I_c$ for the Azores, Siberian, and North American maxima and for the Icelandic and Aleutian minima were obtained from both reanalysis data and model calculations. Moreover, model estimates of the intensity of these ACAs are in good agreement with those obtained from reanalysis data. However, the somewhat lower mean winter intensity of the Aleutian Low, as estimated by the CMIP5 model ensemble, is in the range of standard deviations relative to those obtained from the reanalysis data and from calculations with the CMIP6 model ensemble. At the same time, for the Hawaiian and Arctic maxima, the estimates of the mean winter intensity $I_c$ based on model calculations were found to be statistically insignificant (at least at the 95% significance level), as well as according to the reanalysis data for the period 1981-2005. As for the winter Arctic High, it is characterized by the maximum interannual variability of intensity.

The corresponding estimates of the annual mean ACA intensity $I_c$ in Table 4b according to model calculations are significant and are in good agreement with estimates from reanalysis data for the subtropical Azores and Hawaiian Highs and for the Asian Low. At the same time, for the areas of the Arctic High and the Aleutian, Icelandic and North American Lows, reanalysis data yielded mean summer sea level pressure anomalies of the opposite sign, although insignificant for the Arctic High and the Icelandic and North American Lows. It is worth noting that for the period 1981-2005, in contrast to the assessment from reanalysis data, the model calculations yielded positive (though insignificant) mean summer pressure anomalies in the region of the Arctic High. For the North American Low, both model calculations and reanalysis data showed insignificant estimates of mean summer intensity. The results obtained indicate that ACAs, whose intensity is determined according to (1), are generally more significant in winter seasons than in summer seasons, despite the stronger interannual variability (characterized by the interannual standard deviation).

In order to obtain more reliable estimates of possible changes in ACA regimes under expected climate changes, an adequate reproduction of modern ACA regimes in model calculations is a necessary condition. According to the results obtained, the reproduction of the ACA in both winter and summer according to calculations with the ensemble of CMIP6 climate models is generally significantly better than with the ensemble of CMIP5 climate models. At the same time, as shown in Table 4, for some ACAs the agreement with reanalysis-based estimates is better for the CMIP6 climate models, while for other ACAs it is better for the CMIP5 climate models.

Table 5 shows the intermodel standard deviations for the ACA intensity in the NH in winter and summer for "all" and "best" CMIP5 and CMIP6 models. According to Table 5, the inter-model standard deviations for the ACA intensity in summer for CMIP6 models than for CMIP5 models, for both "all" and "best" models. The opposite was noted only for the Arctic High in summer. For the winter season, a similar tendency was noted for the Arctic, Hawaiian, Siberian, and North American Highs. For the Azores High and the Icelandic Low, this occurs only when "all" models are considered, but not the "best" ones. The opposite tendency was found for the Aleutian and Icelandic Lows and Azores High, and only for the 'best' models.

Table 5. Inter-model standard deviations for the intensity of ACAs in NH in winter and summer according to all and "best" CMIP5 and CMIP6 climate models (in square brackets) for the base period 1981-2005.

|  | inter-model standard deviations | | | |
| :---: | :---: | :---: | :---: | :---: |
| ACAs Northern Hemisphere | winter (December-January-February) | | summer (June-July-August) | |
|  | CMIP5 models | CMIP6 models | CMIP5 models | CMIP6 models |
| Arctic High | 4.0 [2.3] | 2.8 [2.0] | 2.0 [1.4] | 2.1 [1.4] |
| Greenland High | 4.3 [1.6] | 3.3 [2.2] | 3.0 [1.3] | 2.3 [1.5] |
| Islandic Low | 2.0 [1.2] | 1.9 [2.0] | 1.7 [1.2] | 1.5 [0.9] |
| Aleutian Low | 2.3 [1.8] | 2.4 [2.3] | 1.8 [2.0] | 1.7 [1.5] |
| North American High | 2.1 [1.2] | 1.1 [0.8] |  |  |
| Siberian High | 1.3 [1.2] | 1.1 [1.0] |  |  |
| Azores High | 2.1 [1.2] | 1.5 [1.3] | 0.9 [0.7] | 0.9 [0.4] |
| Hawaiian High | 1.7 [1.0] | 1.6 [0.7] | 1.6 [1.3] | 1.4 [1.2] |
| North American Low |  |  | 2.0 [1.7] | 1.9 [1.1] |
| Asian Low |  |  | 1.6 [0.7] | 1.3 [0.7] |

### 3.2.2. Southern Hemisphere

Table 6 presents quantitative estimates of the intensity $I_c$ and their standard deviations for the main ACAs in the SH for the winter and summer seasons, estimated for the ERA5 reanalysis and for the ensemble means according to simulations with "all" and "best" CMIP5 and CMIP6 models for the base period 1981-2005.

Table 6. Same as Table 4, but for the SH.

(a)

| ACAs Southern Hemisphere | winter (June-July-August) | | |
|---|---|---|---|
| | реанализ | модели CMIP5 | модели CMIP6 |
| Australian High | 4.7 (±0.7) | 6.6 (±0.6) [6.6 (±0.6)] | 6.8 (±0.6) [6.8 (±0.6)] |
| South American High | 6.0 (±0.9) | 8.9 (±0.9) [9.0 (±1.0)] | 8.5 (±1.0) [8.5 (±0.9)] |
| South African High | 10.5 (±0.7) | 12.2 (±0.9) [12.2 (±0.8)] | 12.1 (±0.8) [12.0 (±0.8)] |
| South Pacific High | 7.2 (±1.5) | 7.9 (±1.5) [8.5 (±1.3)] | 8.5 (±1.6) [8.1 (±1.5)] |
| South Atlantic High | 9.2 (±0.9) | 10.9 (±1.0) [11.2 (±1.0)] | 11.4 (±1.0) [10.9 (±1.0)] |
| South Indian High | 11.3 (±1.2) | 12.2 (±1.2) [13.2 (±1.1)] | 13.2 (±1.1) [13.0 (±1.0)] |
| South Atlantic Low | -15.8 (±0.9) | -13.6 (±1.7) [-12.8 (±1.6)] | -13.6 (±1.6) [-14.3 (±1.6)] |
| Indian Ocean Low | -17.8 (±1.3) | -16.3 (±2.0) [-16.3 (±1.8)] | -16.6 (±1.8) [-16.9 (±1.8)] |
| South Pacific Low | -20.7 (±4.0) | -17.7 (±3.7) [-17.3 (±3.8)] | -17.3 (±3.9) [-17.9 (±3.7)] |
| Antarctic High | 2.5 (±3.6) | 1.1 (±4.6) [-6.5 (±4.3)] | -1.3 (±4.3) [-0.9 (±4.7)] |

(b)

| ACAs Southern Hemisphere | summer (December-January-February) | | |
|---|---|---|---|
| | reanalysis | CMIP5 models | CMIP6 models |
| Australian Low | 0.5 (±0.8) | 0.7 (±0.7) [0.3 (±0.7)] | 0.5 (±0.7) [0.3 (±0.7)] |
| South American Low | 2.6 (±0.5) | 1.7 (±0.6) [1.8 (±0.6)] | 1.5 (±0.6) [2.1 (±0.5)] |
| South African Low | 4.4 (±0.4) | 4.1 (±0.6) [4.2 (±0.6)] | 3.9 (±0.6) [4.3 (±0.6)] |
| South Pacific High | 9.8 (±1.0) | 8.7 (±0.9) [9.3 (±1.0)] | 9.6 (±0.9) [9.7 (±0.9)] |
| South Atlantic High | 7.9 (±0.6) | 7.8 (±0.9) [8.1 (±0.9)] | 8.3 (±0.7) [8.2 (±0.7)] |
| South Indian High | 8.7 (±0.7) | 8.3 (±0.9) [8.6 (±0.8)] | 8.8 (±0.7) [8.7 (±0.7)] |
| South Atlantic Low | -13.5 (±1.0) | -12.2 (±1.3) [-12.6 (±1.4)] | -12.7 (±1.1) [-13.1 (±1.1)] |
| Indian Ocean Low | -12.4 (±1.1) | -11.3 (±1.3) [-11.6 (±1.2)] | -12.0 (±1.1) [-12.3 (±1.1)] |
| South Pacific Low | -18.3 (±2.8) | -15.8 (±2.9) [-16.4 (±2.9)] | -17.6 (±2.7) [-17.5 (±2.6)] |
| Antarctic High | -1.7 (±3.4) | -1.4 (±4.1) [-6.6 (±4.2)] | -7.4 (±3.5) [-5.8 (±3.5)] |

In the SH, significant (at the 99% or higher level of statistical significance) estimates of the intensity $I_c$ averaged for the winter season were obtained from both reanalysis data and model simulations for all ACAs, except for the Antarctic High (Table 6). Model estimates of the intensity of these ACAs generally agree well with those obtained from reanalysis data. The corresponding estimates of the average ACAs intensity $Ic$ in summer in the SH are in Table 6b by model simulations are significant and agree quite well with estimates from reanalysis data for all ACAs over the oceans. The results obtained indicate that the ACAs are generally manifested more significantly in the winter seasons than in the summer, although the interannual variability (characterized by the standard deviation) of the ACA intensity is stronger in summer. In particular, the cyclonic ACAs over South America, Africa, and Australia are poorly identified in summer in the SH, in contrast to the clearer features of the anticyclonic ACAs over these continents in the winter. The Antarctic anticyclonic ACA is also poorly diagnosed in summer.

In the SH, according to both reanalysis data and model simulations, the average intensity of the subtropical anticyclonic ACA over the Pacific Ocean for the summer seasons is greater, and the interannual variability is less than for the winter seasons. At the same time, for the subtropical anticyclonic ACAs over the Atlantic and Indian Oceans in the SH, both reanalysis data and model simulations show that the mean intensity and interannual variability of the subtropical anticyclonic ACAs are lower for the summer seasons than for the winter seasons.

For subpolar cyclonic ACAs over the Atlantic and Pacific oceans in the SH, both reanalysis data and model simulations show that the mean intensity is generally greater in winter seasons than in summer, with the exception of the South Pacific Low when ensemble of "all" CMIP6 models is used. The interannual variability of the intensity of subpolar cyclonic ACAs in the SH is less in summer than in winter. At the same time, the reanalysis data show a slight increase in the interannual variability of the intensity of the South Atlantic Low in summer compared to winter.

According to the results obtained, the reproduction of the ACAs in the SH in both winter and summer is generally significantly better for CMIP6 climate models than for CMIP5 climate models when the ensembles of "all" models are compared. For "best" models, the agreement with reanalysis-based estimates for some ACAs is better for the CMIP6 ensemble, while for other ACAs it is better for the CMIP5 ensemble.

Table 7 shows inter-model standard deviations for the intensity of ACAs in the SH in summer and winter according to simulations with ensembles of "all" and "best" CMIP5 and CMIP6 climate models. Inter-model standard deviations for the ACA intensity in winter according to simulations with the CMIP6 ensemble of models are generally smaller than with the CMIP5 ensemble of models (Table 7). The opposite was found only for the Australian maximum for the ensemble of selected models (not shown). The differences in the inter-model standard deviations for summer ACAs intensity from simulations with the CMIP6 and CMIP5 model ensembles vary significantly among ACAs.

Table 7. Same as Table 5, but for the SH.

| ACAs Southern Hemisphere | inter-model standard deviations | | | |
|---|---|---|---|---|
| | summer (December-January-February) | | winter (June-July-August) | |
| | CMIP5 models | CMIP6 models | CMIP5 models | CMIP6 models |
| South American Low | 1.4 [0.9] | 1.3 [0.8] | | |
| South African Low | 1.0 [0.3] | 0.8 [0.4] | | |
| Australian Low | 1.2 [0.8] | 0.7 [0.5] | | |
| South Pacific High | 1.5 [1.0] | 1.2 [1.1] | 1.5 [0.8] | 1.4 [0.8] |
| South Atlantic High | 1.2 [1.2] | 1.3 [0.9] | 1.4 [0.9] | 1.3 [0.8] |
| South Indian High | 0.9 [0.5] | 0.8 [0.6] | 1.6 [0.8] | 1.4 [0.4] |
| South Atlantic Low | 1.4 [0.8] | 1.5 [0.8] | 2.4 [1.4] | 2.2 [1.1] |
| Indian Ocean Low | 1.7 [1.2] | 1.3 [0.6] | 3.7 [2.0] | 2.6 [0.9] |
| South Pacific Low | 2.9 [1.7] | 3.0 [2.4] | 4.1 [2.3] | 3.1 [1.9] |
| Antarctic High | 12.2 [5.3] | 6.0 [5.3] | 18.8 [7.3] | 11.0 [6.9] |

### 3.3. Ensemble model estimates of expected changes in ACA characteristics in the 21st century

### 3.3.1. Northern Hemisphere

This section presents estimates of possible changes in the ACAs intensity in the NH from simulations with the CMIP5 and CMIP6 ensembles of "all" and 'best' climate models with the RCP8.5 and the SSP5-8.5 scenarios, respectively. Figure 7 presents estimates of the relative (normalized to the standard deviation for the base period 1981-2005) changes in the intensity of $I_c'$ of the Azores, Siberian and North American High and the Aleutian and Icelandic Lows in winter during the 21st century. For these ACAs, the best agreement of intensity by model simulations and reanalysis data was obtained for the base period 1981-2005.

According to Fig. 7, the changes in the intensity of different ACAs differ significantly, including in sign. The range of variability for different ACAs generally increases in the 21st century. For the winter Azores High the intensity increases significantly by the end of the 21st century under the RCP8.5 scenario. The winter Aleutian Low intensifies also substantially, especially by CMIP5 model simulations. At the same time, a decrease in the intensity of the Siberian and North American Highs and Icelandic Low in winter is found for the 21st century, the most significant for the North American High and the least significant for the Siberian High. The intensity of the winter Azores High increases by the end of the 21st century according to simulations with both CMIP6 and CMIP5 models, more substantially for CMIP5 models (Fig. 7a). Distinct projected changes were also found for the Icelandic Low in winter (Fig. 7e), whose normalized intensity changes insignificantly according to the ensemble mean of the CMIP6 models and decreases according to those of the CMIP5 models.

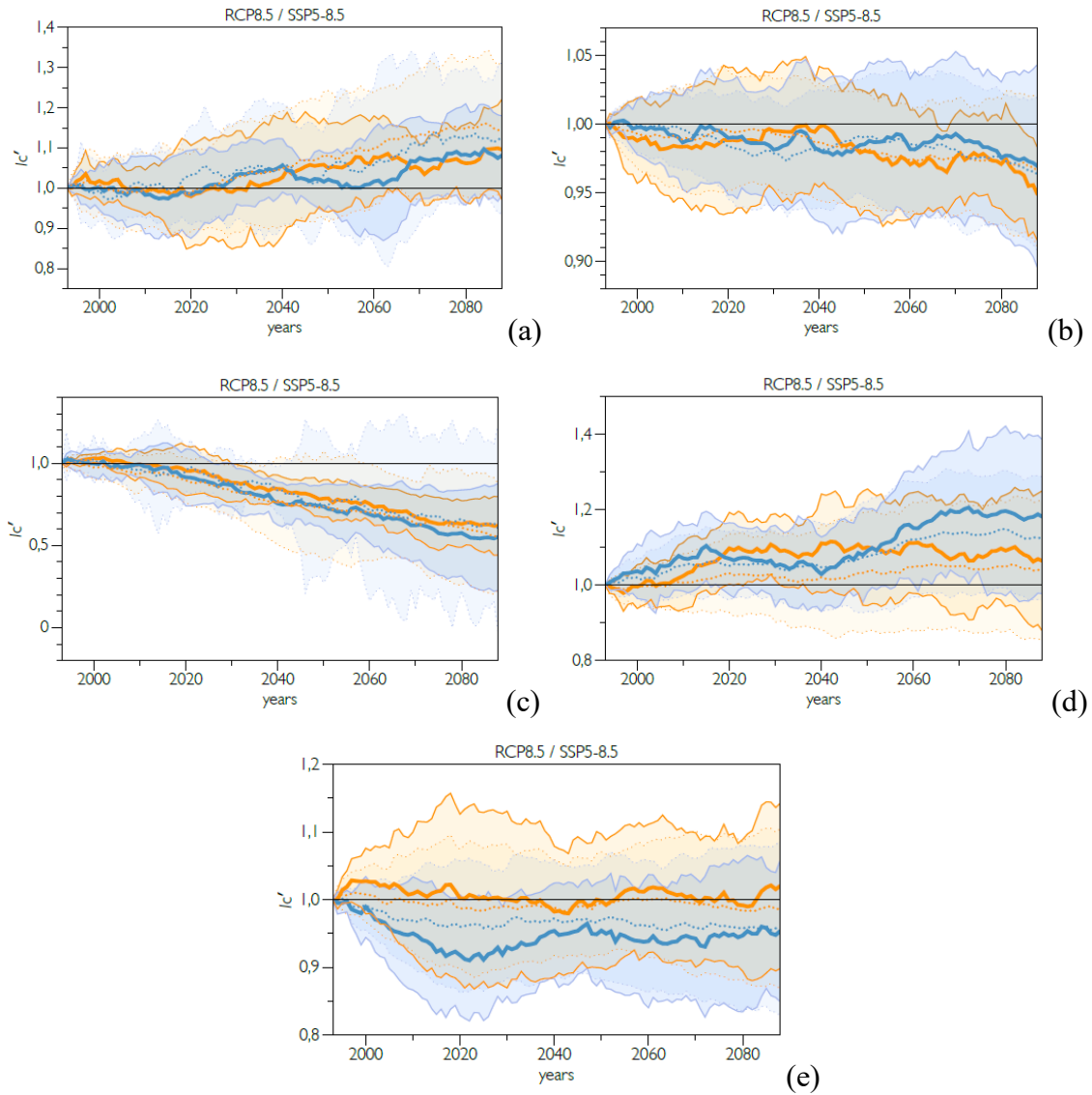

Figure 7. Changes in the intensity $I_c'$ (normalized to the mean for the 1981–2005 base period) of the main winter ACAs in the NH: (a) Azores High, (b) Siberian High, (c) North American High, (d) Aleutian Low, (e) Icelandic Low. 25-year moving averages are shown for with ensembles of the CMIP5 models under the RCP8.5 scenario (blue color) and the CMIP6 models under the SSP5-8.5 scenario (orange color), both merged with the corresponding historical scenario. Thick lines correspond to the ensemble mean; thin lines characterize the ranges (with shading) of the standard inter-model deviations. Solid lines represent for "best" models, dashed lines represent "all" models. Note, that $I_c'>1$ corresponds to ACA strengthening (both for lows and highs), and $I_c'<1$ corresponds to ACA weakening.

Similar to Fig. 7 for the winter ACAs in NH, Fig. 8 shows the corresponding changes in the normalized intensity of the Azores and Hawaiian Highs and the Asian Low in the summer seasons. A significant weakening was obtained for the summer Asian Low. An insignificant intensification was noted for the Hawaiian High. For the Azores High in the summer seasons, in comparison to the winter seasons, less significant changes were found in the 21st century (Fig. 8a). In general, the CMIP6 and CMIP5 models are in agreement on the projected changes of ACAs intensity in the summer seasons (Fig. 8).

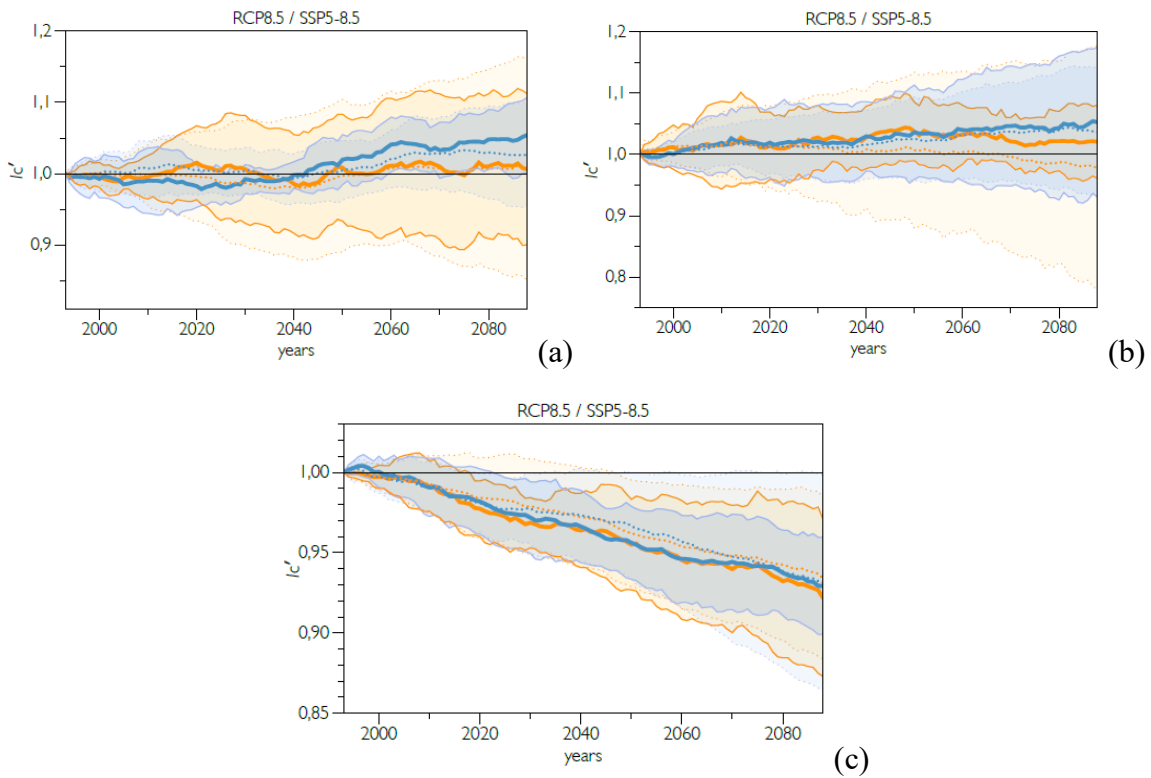

Figure 8. Same as Figure 7, but for the summer season (JJA) for (a) Azores High, (b) Hawaiian High, and (c) Asian Low.

### 3.3.2. Southern Hemisphere

Figure 9 presents estimates of changes in the 21st century in the relative intensity $I_c'$ (normalized to the standard deviation for the 1981-2005 base period) of the South Pacific, South Atlantic, Indian Ocean, South American, South African and Australian maxima, as well as the South Pacific, South Atlantic and Indian Ocean minima for the winter seasons. In general, the tendencies of significant increasing intensity for winter ACAs common to the CMIP6 and CMIP5 model ensembles appear in the SH in the 21st century with the exception of the South American and South Atlantic Highs, which show insignificant changes.

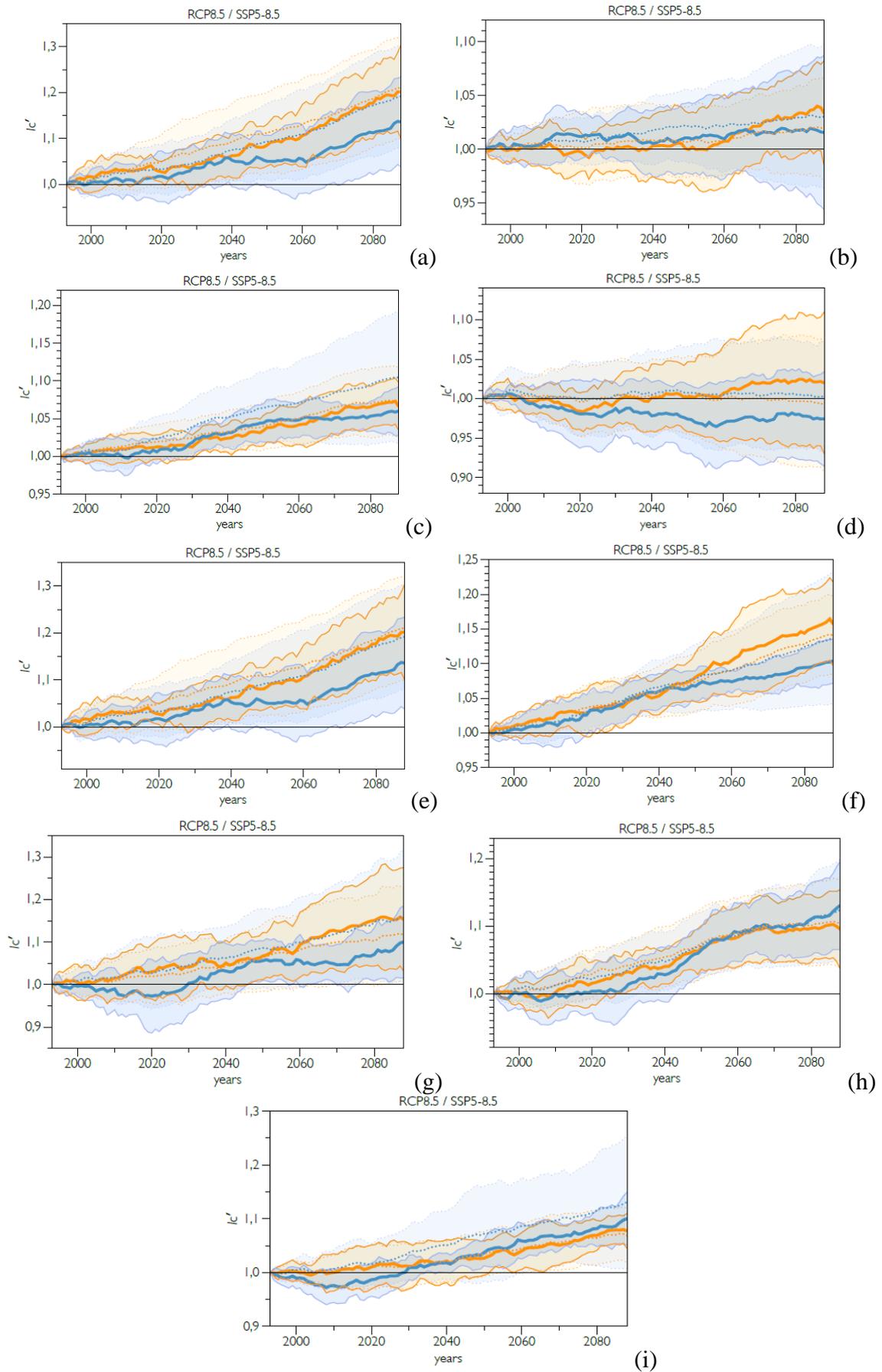

Figure 9. Same as Fig. 7, but for the SH for the winter season (JJA) for (a) South Pacific High, (b) South Atlantic High, (c) Indian Ocean High, (d) South American High, (e) South African High, (f) Australian High, (g) South Pacific Low, (h) South Atlantic Low, (i) Indian Ocean Low.

Figure 10 presents the corresponding estimates of changes in the relative ACAs intensity $I_c'$ (normalized to the standard deviation for the 1981-2005 base period) in the SH during the 21st century for the summer seasons (DJF): South Pacific, South Atlantic and Indian Ocean maxima and South Pacific, South Atlantic and Indian Ocean Lows. In general, in summer, as in winter, there is a general strengthening in the 21st century of the analyzed ACAs in the SH, more significant for South Pacific ACAs. The intensification of the South Pacific High is remarkably weaker according to simulations with the CMIP6 ensemble than according to simulations with the CMIP5 ensemble in summer seasons in contrast with winter seasons. The observed relative intensification of the corresponding ACAs in summer is generally less than in winter. The largest relative increase in intensity was obtained for the summer Pacific ACAs – subtropical maximum and subpolar minimum.

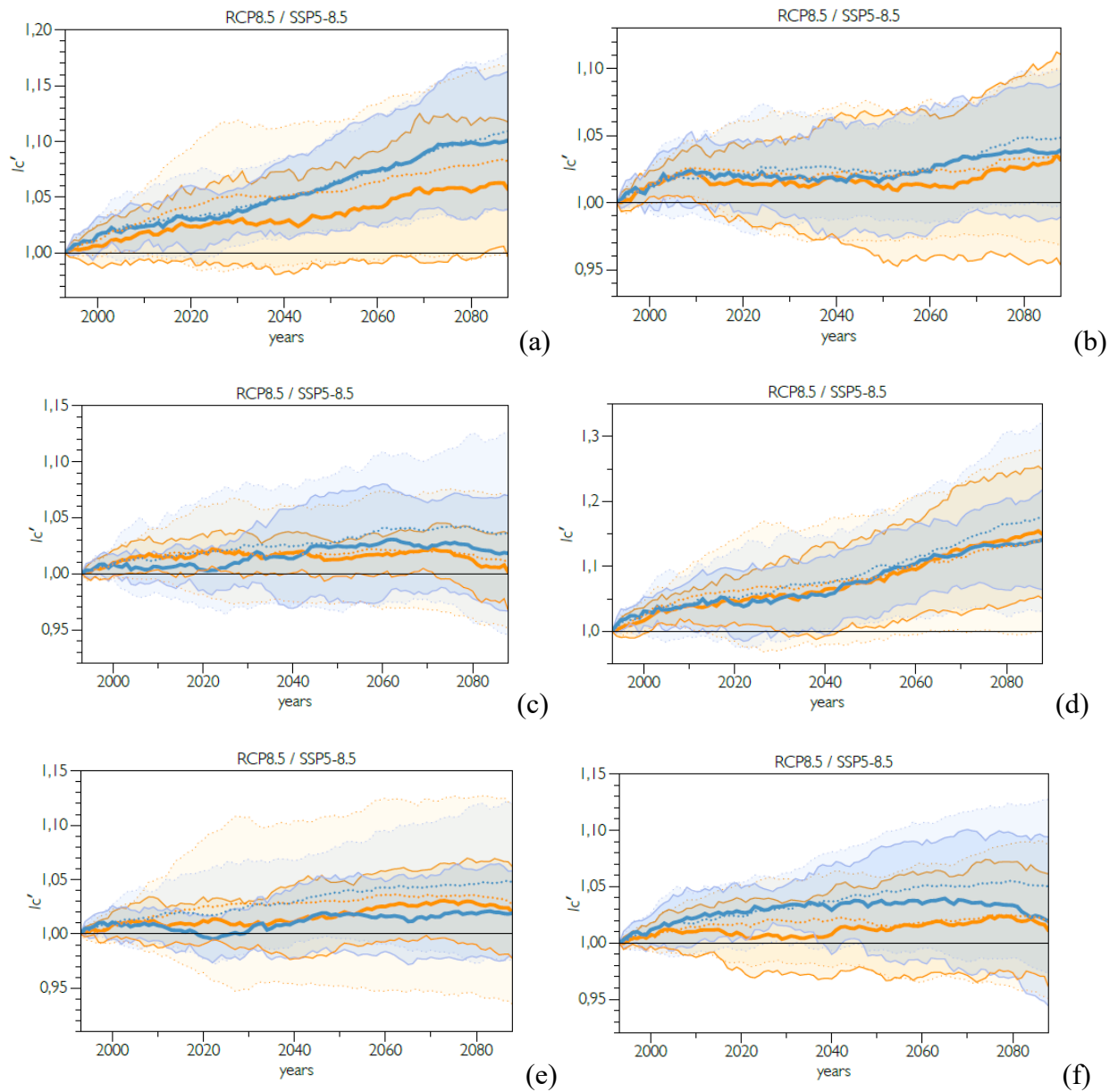

Figure 10. Same as Fig. 7, but for the SH for the summer season (DJF) for (a) South Pacific High, (b) South Atlantic High, (c) Indian Ocean High, (d) South Pacific Low, (e) South Atlantic Low, (f) Indian Ocean Low.

## 4. Conclusion

The analysis of possible changes in the ACA features by simulations with the CMIP5 and CMIP6 ensembles of climate models revealed ambiguous trends of changes for different ACAs in the NH, in particular under the scenarios of anthropogenic forcings RCP8.5 and SSP5-8.5 in the 21st century. The obtained estimates have different levels of statistical significance. With the CMIP5 and CMIP6 model ensembles, the most consistent estimates were obtained for the weakening trends of the winter North American High and summer Asian Low. For the winter Siberian High, the weakening trend was found to be more significant by simulations with the CMIP6 ensemble of climate model. In the detection of ACAs and the assessment of trends in their intensity, a number of seasonal and regional peculiarities were noted. In particular, for the 1981-2005 base period, according to both reanalysis data and model simulations, the average intensity of the subtropical anticyclonic ACAs over the Atlantic and Pacific Oceans is generally more significantly determined for the summer seasons than for the winter. The opposite is observed for the average intensity of subpolar cyclonic ACAs over the Atlantic and Pacific Oceans - their intensity is generally more significantly determined for the winter seasons than for the summer. This is manifested despite the interannual variability of ACA intensity (characterized by the standard deviation) is generally greater in the winter seasons.

In the SH, for the subtropical anticyclonic ACAs over the Atlantic and Indian Oceans, both reanalysis data and model simulations show that the mean intensity and interannual variability are lower for the summer seasons than for the winter. At the same time, for the subtropical anticyclonic ACA over the Pacific Ocean, the mean intensity is greater for the summer seasons than for the winter, while the interannual variability is less. For the subpolar cyclonic ACAs over the Atlantic and Pacific Oceans in the SH, both reanalysis data and model simulations show that the mean intensity is generally greater in winter than in summer, with the exception of the South Pacific Low when "all" CMIP6 models are considered. The interannual variability of the intensity of subpolar cyclonic ACAs in the SH is less in summer than in winter. The reanalysis data show a slight increase in the interannual variability of the intensity of the South Atlantic Low in summer compared to winter. According to the obtained ensemble model estimates, particularly according to the CMIP6 models simulations, there is a general strengthening of the analyzed ACAs in the SH in summer and winter with warming in the 21st century. The noted relative increase in summer ACAs intensity is generally less than in winter. The largest increase in relative intensity was found for the summer Pacific ACAs, i.e., the subtropical maximum and the subpolar minimum.

This study was carried out within the framework of the RSF project No. 24-17-00211 with the use of results obtained within the RSF project No. 24-17-00357. The additional support for analysis of projected changes of ACA intensity have been provided by the state assignment of the A. M. Obukhov Institute of Atmospheric Physics RAS (FMWR-2022-0014).